\documentclass[crcready]{iosart2c}

\usepackage[T1]{fontenc}
\usepackage{times}%

\usepackage{amsmath}
\usepackage{dcolumn}
\usepackage{graphics}
\usepackage{graphicx}

\newcolumntype{d}[1]{D{.}{.}{#1}}

\firstpage{1} \lastpage{6} \volume{1} \pubyear{2012}

\begin{document}
\begin{frontmatter}                           

%
\title{TheSoz: A SKOS Representation of the Thesaurus for the Social Sciences}

\runningtitle{TheSoz: A SKOS based Thesaurus for the Social Sciences}

\review{Pascal Hitzler, Kno.e.sis Center, Wright State University, Dayton, OH, USA; Krzysztof Janowicz, University of California, Santa Barbara, USA}{Ivan Herman, W3C; Christophe Gu\'eret, Vrije Universiteit Amsterdam, The Netherlands; Danh Le Phuoc, DERI Galway, Ireland}{}

\author[A]{\fnms{Benjamin} \snm{Zapilko}\thanks{Corresponding author. E-mail: benjamin.zapilko@gesis.org.}},
\author[A]{\fnms{Johann} \snm{Schaible}}
\author[A]{\fnms{Philipp} \snm{Mayr}}
and
\author[A]{\fnms{Brigitte} \snm{Mathiak}}
\runningauthor{B. Zapilko et al.}
\address[A]{GESIS - Leibniz Institute for the Social Sciences, Knowledge Technologies for the Social Sciences, Unter Sachsenhausen 6-8, 50667 Cologne, Germany\\
E-mail: \{benjamin.zapilko,johann.schaible,philipp.mayr,brigitte.mathiak\}@gesis.org}

\begin{abstract}
The Thesaurus for the Social Sciences (TheSoz) is a Linked Dataset in SKOS format, which serves as a crucial instrument for information retrieval based on e.g. document indexing or search term recommendation. Thesauri and similar controlled vocabularies build a linking bridge for datasets from the Linked Open Data cloud. In this article the conversion process of the TheSoz to SKOS is described including the analysis of the original dataset and its structure, the mapping to adequate SKOS classes and properties, and the technical conversion. In order to create a semantically full representation of TheSoz in SKOS, extensions based on SKOS-XL had to be defined. These allow the modeling of special relations like compound equivalences and terms with ambiguities. Additionally, mappings to other datasets and the appliance of the TheSoz are presented. Finally, limitations and modeling issues encountered during the creation process are discussed. 
\end{abstract}

\begin{keyword}
Thesaurus\sep Linked Data\sep SKOS\sep Social Sciences
\end{keyword}

\end{frontmatter}

\section{Introduction}
The Thesaurus for the Social Sciences (TheSoz) is a SKOS-based German thesaurus for the domain of the social sciences. It serves as a crucial instrument for indexing documents and research information as well as for search term recommendation. The TheSoz is available in three languages (German, English and French) and contains overall about 12,000 keywords, from which 8,000 are so-called descriptors, i.e. preferred terms for indexing documents, and 4,000 non-descriptors, i.e. non-preferred terms, for which preferred terms are recommended to be used instead. The thesaurus covers all topics and sub-disciplines of the social sciences such as sociology, employment research, pedagogics or political science. Additionally terminology from associated and related disciplines like economics is included in order to support an accurate and adequate indexing process of interdisciplinary, praxis-oriented and multicultural documents. The thesaurus is owned and maintained by GESIS - Leibniz Institute for the Social Sciences\footnote{http://www.gesis.org/en/home/}, an infrastructure organization in Germany, which provides research-based infrastructure services for the social sciences. Although TheSoz is specific to the social sciences, most of the terms are used in the colloquial German language, especially in classic media and political texts. 

First attempts \cite{17} for representing the TheSoz in SKOS (Simple Knowledge Organization System)\footnote{http://www.w3.org/2004/02/skos/} format have been made in 2009 after SKOS has been announced as a standard by the W3C. A lot of organizations and libraries have begun bringing their thesauri and vocabularies to the web in SKOS format since then.

The SKOS version of the TheSoz is based on an intellectually maintained database. It is currently available in version 0.92 via a SPARQL endpoint\footnote{http://lod.gesis.org/thesoz/sparql} as well as for download\footnote{http://www.gesis.org/en/services/research/thesauri-und-klassifikationen/social-science-thesaurus/} in RDF/XML and RDF/Turtle under a Creative Commons Licence\footnote{http://creativecommons.org/licenses/by-nc-nd/3.0/}. Currently, it consists of 421,083 triples. Each URI is HTTP dereferenceable, which has been enabled by a representation via the Pubby Linked Data Frontend \cite{12}. This HTML representation is accessible at http://lod.gesis.org/thesoz/. A user-friendly RDFa-supported interface is planned for the future.

TheSoz also provides crosslinks to other thesauri: the STW\footnote{http://zbw.eu/stw/versions/latest/about.en.html}, the equivalent German thesaurus for economics, the AGROVOC\footnote{http://aims.fao.org/website/AGROVOC-Thesaurus/sub} thesaurus for the agricultural domain as well as to DBpedia\footnote{http://dbpedia.org/}. Although the subject matters of the connected thesauri are similar in some cases, e.g. between TheSoz and STW, the terms and concepts used, are quite differently constructed. Links between thesauri build a relevant bridge for the connection between different Linked Data sources.

In section 2 of this article the conversion process of the TheSoz to SKOS is presented including the definition of extensions based on SKOS-XL. Furthermore, the use of established vocabularies and links to other thesauri are described. In section 3 the appliance of the TheSoz for information retrieval purposes is presented. Typical knowledge modeling patterns occurring by the use of SKOS and SKOS-XL are discussed in section 4. Section 5 concludes and presents an outlook on the future work regarding the TheSoz.

\section{Conversion of the TheSoz to SKOS}
The SKOS version of the TheSoz is created from the original source, which is maintained and stored in a customized data management system at GESIS. Content updates for the thesaurus are released regularly every month, followed directly by an update of the SKOS version. The transformation process of the thesaurus data into SKOS format has been split up into three steps and has thereby followed the structured method introduced in \cite{2}, which consists of the following steps: (1) analysis of the structure, the extent and the complexity of the thesaurus, including contained terms and relations between terms, (2) a mapping of all detected terms and relations to adequate SKOS classes and properties and (3) the technical conversion of the thesaurus according to the defined mapping. This method \cite{2} aims to ensure the quality and utility of the resulted conversion regarding its interoperability and completeness. All three steps of the method are applied once on the thesaurus in order to generate an initial SKOS version. The third step, the technical conversion and creation of the target output is conducted regularly, which depends on updates of the content of the thesaurus.

\subsection{Thesaurus Creation}
The basis for a transformation of a thesaurus to SKOS format builds a detailed analysis of the thesaurus. Therefore, attention has not only been paid to terms and existing associative and hierarchical relations between them, but also to the general structure and design issues of the thesaurus, e.g. the existence of an additional classification system or how far the thesaurus conforms to established ISO norms\footnote{ISO norms for thesauri 2788, 5964 and 25964-1}. Relationships between the 8,000 TheSoz descriptors are expressed as broader, narrower or related terms. There are also "use instead" and "use combination" relations and their counterparts ("used for" and "used for combination") between descriptors and non-descriptors. Additionally a classification hierarchy is provided and each thesaurus term is assigned to one or more classification notations.

\begin{table*}[htb]
\caption{Overview on personal extensions defined for the TheSoz.} \label{t1}
\begin{tabular}{p{.3\textwidth}p{.6\textwidth}}
\hline
Extension&Description\\
\hline
thesoz:Descriptor&Descriptors of the TheSoz, which are defined as subclasses of "skos:Concept".\\
thesoz:Classification&Notation of the classification hierarchy of the TheSoz, which is defined as a subclass of "skos:Concept".\\
thesoz:EquivalenceRelationship&An equivalence relationship between two terms, where the terms are assigned via "thesoz:use" and "thesoz:usedFor" properties. This is a subclass of "skosxl:Label".\\
thesoz:CompoundEquivalence&A compound equivalence between terms. For constructing "use combination" and "used for combination" relations between terms. The non-preferred term is assigned by the "thesoz:compoundNonPreferrdTerm" property, the preferred terms by the "thesoz:preferredTermComponent" property. This is a subclass of "skosxl:Label".\\
thesoz:use&Use relation, which is defined as a subproperty of "skosxl:labelRelation".\\
thesoz:usedFor&Used for relation, which is defined as a subproperty of "skosxl:labelRelation".\\
thesoz:preferredTermComponent&A preferred term as a component for a "use combination" and "used for combination" relation. This property is defined as a subproperty of "skosxl:labelRelation".\\
thesoz:compoundNonPreferredTerm&The non-preferred term as a component for a "use combination" and "used for combination" relation. This property is defined as a subproperty of "skosxl:labelRelation".\\
thesoz:isPartOfEquivalenceRelationship&Relation from a term to the class "thesoz:EquivalenceRelationship".\\
thesoz:isPartOfCompoundEquivalence&Relation from a term to the class "thesoz: CompoundEquivalence".\\
thesoz:hasTranslation&Relation between different languages of a term, which is defined as a subproperty of "skosxl:labelRelation".\\
thesoz:isTranslationOf&Inverse property of "thesoz:hasTranslation".\\
\hline
\end{tabular}
\end{table*}

For most of the thesaurus terms and relations adequate SKOS properties and classes can easily be identified because of a broad compatibility of the TheSoz to the standard norms for thesauri. Problems have been observed, when mapping special data items and relations, which are not compliant to thesauri standards like the AD (alternative non-descriptor) terms of the TheSoz. An alternative non-descriptor is used for indicating ambiguities. Therefore, terms of this type describe generic and ambiguous terms, which have different meanings in specialized sub-contexts. This is expressed through the use of multiple "use instead" and/or "use combination" relations for one single term at the same time. There are about 200 of such AD terms in the TheSoz.

For example, the term "committee", which is classified as an AD term, holds "use" relations to the preferred terms "working group", "parliamentary committee", "Wirtschaftsausschuss" and "advisory panel" at the same time. Additionally, it contains a "use combination" relation to the combined use of the terms "product" and "quality". In this case, it means that the term "committee" is in its semantics such general and ambiguous, that it is recommended to use a more precise term to describe the intended semantics.

But, because SKOS is based on RDF it is possible to define own relations without greater effort. Therefore, a precise mapping to SKOS has been more complex than initially thought \cite{17}, where only simple relations of the TheSoz have been modeled in SKOS. In order to obey the concept-based structure of SKOS, but without losing relevant relations between preferred and non-preferred terms, classes and properties of SKOS-XL (SKOS eXtension for Labels)\footnote{http://www.w3.org/TR/skos-reference/skos-xl.html} have been used. SKOS-XL has also been used for the representation of the EUROVOC\footnote{http://eurovoc.europa.eu/} thesaurus \cite{13}. Properties of SKOS-XL have been developed explicitly for the representation of lexical issues and provide the possibility to model relations between lexical terms inside one SKOS concept. Because of the interconnection of terms in the TheSoz descriptors are represented as "skos:Concept", but each of the terms is additionally modeled separately as "skosxl:Label". The property "skosxl:labelRelation" allows the definition of extensions such as typical equivalence relationships like "use" or compound equivalence relationships like "use combination". Table 1 provides an overview on the personal classes and properties defined for the TheSoz.

Thus, the term "university ranking" gets the property "thesoz:compoundNonPreferredTerm" from a class of the type "thesoz:CompoundEquivalence". The two other preferred terms ("university" and "ranking"), which should be used instead, are assigned by the property "thesoz:preferredTermComponent" from the same class. This construct indicates that both preferred terms have to be used together instead of the other term (see Figure 1).

\begin{figure}
\includegraphics[width=75mm]{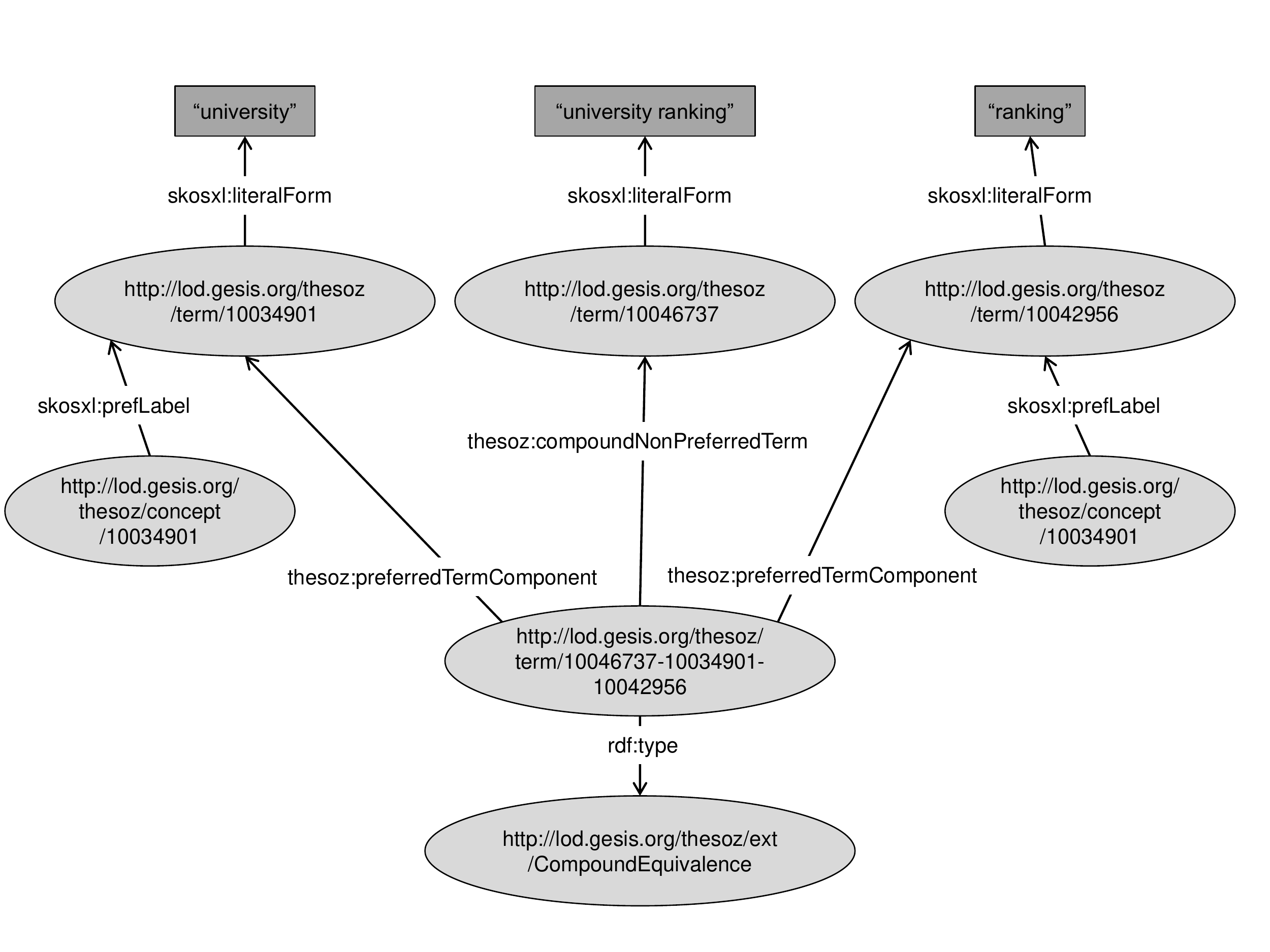}
\caption{Representation of a "use combination" relation in the TheSoz.}
\end{figure}

This modeling approach allows a consistent representation of the alternative non-descriptors of the TheSoz. Thus, the AD term "committee" and its relationships to other terms can be modeled as depicted in Figure 2. It is assigned by multiple "thesoz:usedFor" properties from different classes of the type "thesoz:EquivalenceRelationship", each of which also holds "thesoz:use" properties to the preferred terms "working group", "parliamentary committee", "Wirtschaftsausschuss" and "advisory panel". Additionally, the "thesoz:compoundNonPreferredTerm" property is assigned by a class of the type "thesoz:CompoundEquivalence", which models the "use combination" relation for the combined use of the terms "product" and "quality". Without SKOS-XL based extensions, the ambiguities of the term would be lost.

\begin{figure}
\includegraphics[width=75mm]{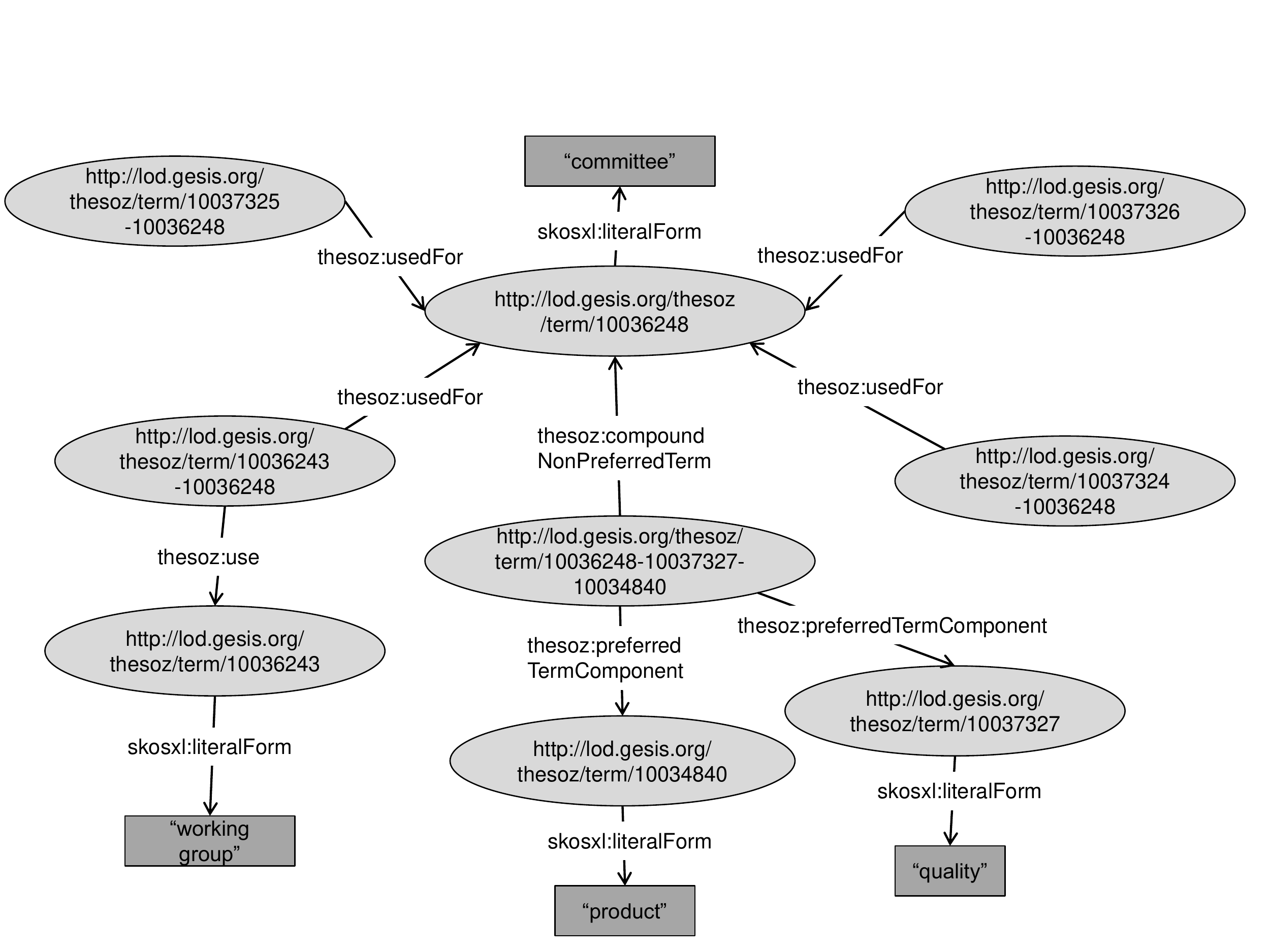}
\caption{Representation of the AD term "committee" in the TheSoz.}
\end{figure}

According to the Linked Data principles, all concepts, terms and classification notations get their own URI, which provides a persistent and unique identification of each elements. This is a very important aspect for re-using and linking TheSoz terms on the web. All URIs are listed in the context path http://lod.gesis.org/thesoz/, which serves as base URI. The URI has been chosen according to naming conventions of web addresses of GESIS and in order to leave room for the publication of further datasets as Linked Data. The namespace of the personal classes and properties is defined at http://lod.gesis.org/thesoz/ext/ and is shortened by the prefix "thesoz".

\subsection{Usage of existing Vocabularies}
Beside the SKOS standard, additional established metadata vocabularies are used to represent the TheSoz. For citation, licensing and provenance purposes properties of Dublin Core, OWL, Creative Commons and the Provenance Vocabulary\footnote{http://trdf.sourceforge.net/provenance/ns.html} have been used. Detailed Information on links to other datasets is represented using the VoiD vocabulary\footnote{http://www.w3.org/TR/void/}. This dataset description is available at http://lod.gesis.org/thesoz/void.ttl. The SKOS-XL extensions of TheSoz are defined in detail using RDF Schema. This ensures further processing and interoperability with other datasets on the web.

\begin{table*}[htb]
\caption{Number of Links from TheSoz to other Datasets.} \label{t1}
\begin{tabular}{ll}
\hline
Dataset&Number and Type of Links\\
\hline
STW&4927 (2844 exact matches, 631 related matches, 1418 broad matches, 34 narrow matches)\\
AGROVOC&846 (840 exact matches, 6 close matches)\\
DBpedia&5024 (all exact matches)\\
\hline
\end{tabular}
\end{table*}

\begin{figure}
\includegraphics[width=75mm]{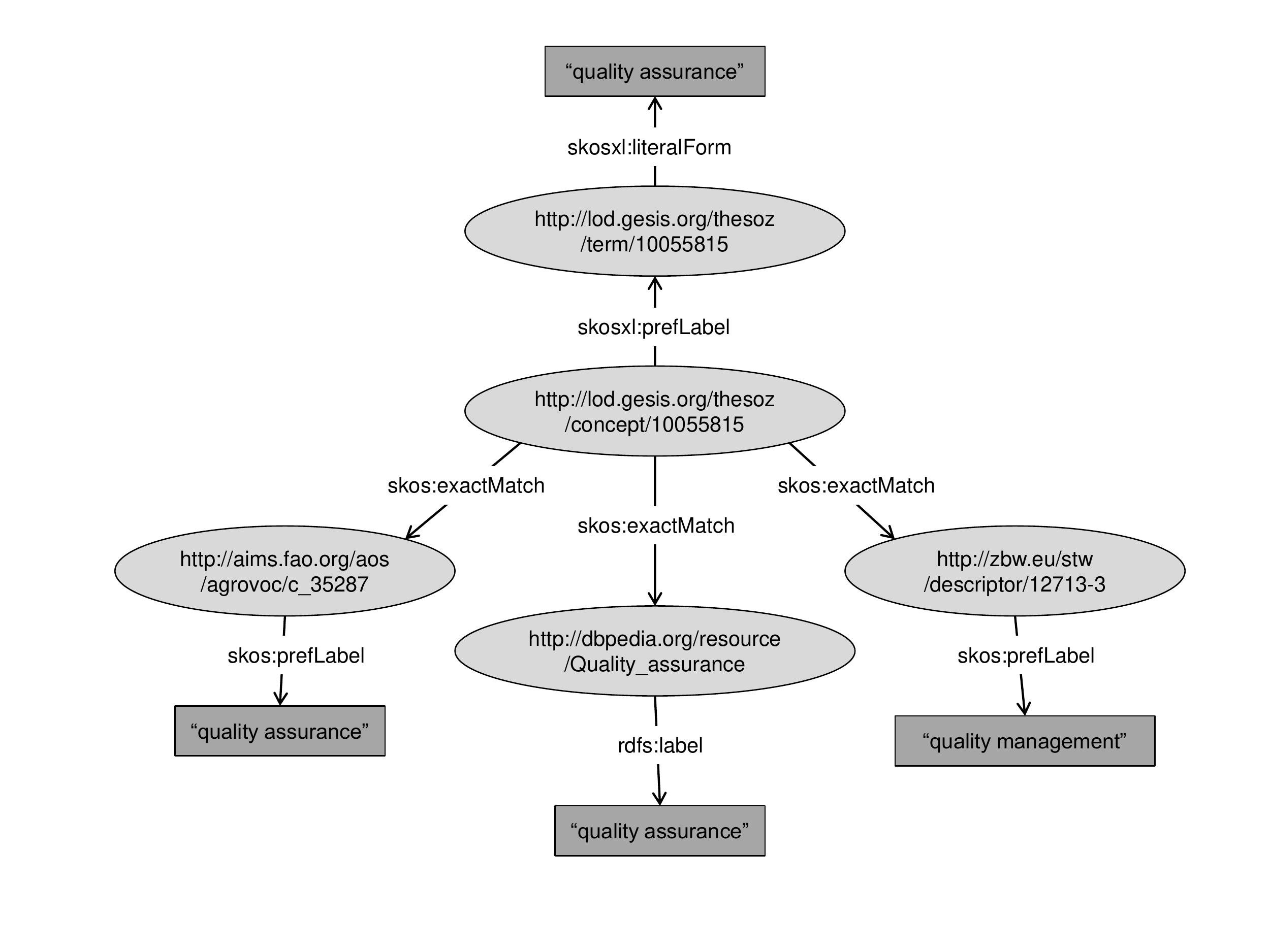}
\caption{Links from the term "quality assurance" to other Datasets.}
\end{figure}

\subsection{Links to other Datasets}
Currently, the SKOS version of the TheSoz holds links to the STW Thesaurus for Economics of the ZBW and the AGROVOC thesaurus of FAO as well as to DBpedia. Such links are important for the use of the thesaurus as a terminology hub during information retrieval, e.g. for search and query expansion. The SKOS mapping properties provide standard relations in order to represent links of SKOS concepts between different concept schemes, i.e. between different datasets. Table 2 provides an overview on number and kind of links from TheSoz to other datasets.

These links have been established differently. While the links to STW have been made manually during a major mapping initiative \cite{8} and have been converted to SKOS via XSL transformation afterwards \cite{9}, the links to AGROVOC and DBpedia have been detected by semi-automatic approaches, which results have been evaluated by domain experts afterwards. The links to AGROVOC have been identified by a distance measure approach using the Levenshtein distance with a threshold of 0.21 \cite{10}. The links to DBpedia have been detected using a standard string similarity algorithm. Figure 3 depicts links from the term "quality assurance" to STW, AGROVOC and DBpedia.

\section{Usage of the TheSoz}
TheSoz is the main tool for indexing research literature in the German speaking social sciences and is applied among other disciplinary information systems in the databases SOLIS (Social Science Literature Information System), SOFIS (Social Science Research Information System), SSOAR (Social Science Open Access Repository) all owned and maintained by GESIS. The thesaurus is used for document retrieval in the information portal sowiport\footnote{http://www.gesis.org/sowiport/en/}, which is visited by about 11,000 unique users per month. This portal uses TheSoz as a terminology hub for search and query expansion, which enables the retrieval of more than 7,000,000 research documents on a semantically very precise level. Other databases of sowiport (e.g. CSA Sociological Abstracts, PAIS International) are indexed with other thesauri than TheSoz. To overcome this heterogeneity links between terms of these thesauri are used to support a meaningful keyword search. The links are used automatically to expand the search terms for the retrieval inside the other non-TheSoz-indexed databases. In \cite{7} the effect of using these mappings for intra- and interdisciplinary search questions has been evaluated in a controlled scenario. The expansion with "exact match"-mappings shows a very positive effect in terms of retrieval precision and recall.

TheSoz and its links to other thesauri are also used for the construction and provision of specialized search term recommendation functionalities, which are outlined and evaluated in \cite{6}. In this case the best recommender has been a system, which starts with recommending TheSoz descriptors and leading to further highly-associated TheSoz descriptors, which have been listed on the basis of a pre-computed co-word analysis.

\section{Modeling Issues}
During the modeling process, several obstacles with the use of SKOS have been observed. Even if a thesaurus meets established ISO norms for thesauri, a conversion to SKOS is not always as trivial as expected \cite{2,11,13}: Relations of the original thesaurus cannot always be modeled adequately in SKOS. For TheSoz especially the representation of compound relationships and compound concepts has been an obstacle. This has also been required by \cite{2,13}, because compound concepts are part of the ISO 2788 standard. Critical modeling issues are discussed in an overview on correspondences between the ISO norms 2788/5964 and SKOS\footnote{http://www.w3.org/TR/skos-primer/\#seccorrespondencesISO} inside the SKOS Primer. For syntactical compositions of terms like compound equivalences, it is suggested to define personal extensions either of "skos:Concept" or "skosxl:Label". For the SKOS version of TheSoz subclasses of both and subproperties of "skosxl:labelRelation" have been defined. \cite{13} applied these relations in a similar way for the EUROVOC thesaurus. \cite{11} has defined a personal construct called "zbwext:useInsteadNote" as a subproperty of "skos:note", which holds information about a "use instead" relation.

When modeling mappings between thesauri in SKOS format, inconsistencies and problems can occur which are caused by idiosyncrasies in thesauri. A reason for inconsistencies can be that given mappings have been defined on term-based thesauri before their conversion to SKOS. Such links cannot always be modeled directly with the SKOS mapping properties. It has to be investigated, if the two terms of a given mapping represent adequate concepts in the corresponding SKOS versions by e.g. being used as "skos:prefLabel" in a concept. Mappings between non-preferred terms cannot directly be modeled in SKOS, because the mapping properties of SKOS can only be applied between concepts. Although ISO 5964 allows relations between non-preferred terms, it is only possible defining and using SKOS-XL extensions.

Domain-specific differences in thesauri can cause conversion problems either. A concept in one thesaurus might correspond to a combination of two concepts in another thesaurus, e.g. the term "Electronic Government" of the TheSoz has originally been mapped to the combination of the terms "Public Administration" and "Internet" of the STW. The mapping properties of SKOS do not allow such single-to-multiple relations (neither for one language nor for multiple languages). 

Transforming existing vocabularies and thesauri to SKOS remains a complex process according to the heterogeneous structure of the involved vocabularies. If semantically more complex relations are required, i.e. "use combination" relations between terms of the same or different vocabularies, personal extensions still have to be defined in order to preserve the relevant information. But, such extensions can lead to incompatibilities to other SKOS datasets or with applications for processing data in SKOS format, e.g. SKOS thesaurus management tools. Therefore extensions should be described using standard classes and properties, e.g. with RDF Schema, so that there is a chance, that such data is processible at least in a minimal way. 

\section{Conclusion and Future Work}
In this paper, the Linked Dataset TheSoz, a widely used thesaurus in the domain of the social sciences, has been presented. It has been brought to SKOS format by the use of SKOS-XL extensions. This has been necessary because of specific and complex relations of the original thesaurus like compound equivalences.

It is planned to add more links to datasets like EUROVOC, the Integrated Authority File (GND)\footnote{http://www.dnb.de/EN/Home/home\_node.html} dataset of the German National Library and more. For some of these tasks it is considered to use link discovery tools like Silk \cite{16} or Amalgame \cite{18}.

TheSoz and other linked vocabularies play a vital role for connecting heterogeneous GESIS datasets like the literature collections of sowiport or the GESIS Data Catalogue\footnote{http://www.gesis.org/en/services/research/data-catalogue/}, which comprises study descriptions of survey data, with each other and with other Linked Datasets from the web. By using Linked Data we aim to provide an integrated view on documents and datasets relevant for social science research.

TheSoz participates as test dataset in the Library Track at the upcoming OAEI 2012\footnote{http://oaei.ontologymatching.org/2012/} together with the STW thesaurus. The existing links between both thesauri serve as reference alignments.

\end{document}